\documentclass[%
 reprint,
 superscriptaddress,
 amsmath,amssymb,
 aps,
floatfix,
]{revtex4-2}

\usepackage{graphicx}% Include figure files
\usepackage{dcolumn}% Align table columns on decimal point
\usepackage{bm}% bold math
\usepackage{hyperref}% add hypertext capabilities
\usepackage{gensymb}
\usepackage{xcolor}
\usepackage{algpseudocode}
\usepackage{algorithm}
\usepackage{lipsum}  
\usepackage{textgreek}

\hypersetup{
    colorlinks=true,
    linkcolor=blue,
    citecolor=blue, 
    urlcolor=black
    }
\begin{document}

\title{Structure of self-generated magnetic fields in laser-solid interaction from proton tomography}

\author{J. Griff-McMahon}
\email{jgriffmc@pppl.gov}
\affiliation{Department of Astrophysical Sciences, Princeton University, Princeton, New Jersey 08544, USA}
\affiliation{Princeton Plasma Physics Laboratory, Princeton, New Jersey 08540, USA}

\author{C. A. Walsh}
\affiliation{Lawrence Livermore National Laboratory, Livermore, California 94550, USA}

\author{V. Valenzuela-Villaseca}
\affiliation{Department of Astrophysical Sciences, Princeton University, Princeton, New Jersey 08544, USA}
\affiliation{Lawrence Livermore National Laboratory, Livermore, California 94550, USA}
\affiliation{Plasma Science and Fusion Center, Massachusetts Institute of Technology, Cambridge, Massachusetts 02139, USA}

\author{S. Malko}
\affiliation{Princeton Plasma Physics Laboratory, Princeton, New Jersey 08540, USA}

\author{B. McCluskey}
\affiliation{Department of Astrophysical Sciences, Princeton University, Princeton, New Jersey 08544, USA}
\affiliation{Princeton Plasma Physics Laboratory, Princeton, New Jersey 08540, USA}

\author{K. Lezhnin}
\affiliation{Princeton Plasma Physics Laboratory, Princeton, New Jersey 08540, USA}

\author{H. Landsberger}
\affiliation{Department of Astrophysical Sciences, Princeton University, Princeton, New Jersey 08544, USA}
\affiliation{Princeton Plasma Physics Laboratory, Princeton, New Jersey 08540, USA}

\author{L. Berzak Hopkins}
\affiliation{Princeton Plasma Physics Laboratory, Princeton, New Jersey 08540, USA}

\author{G. Fiksel}
\affiliation{Center for Ultrafast Optical Science, University of Michigan, Ann Arbor, Michigan 48109, USA}

\author{M. J. Rosenberg}
\affiliation{Laboratory for Laser Energetics, University of Rochester, Rochester, New York 14623, USA}

\author{D. B. Schaeffer}
\affiliation{Department of Physics and Astronomy, University of California Los Angeles, Los Angeles, California 90095, USA}

\author{W. Fox}
\affiliation{Department of Astrophysical Sciences, Princeton University, Princeton, New Jersey 08544, USA}
\affiliation{Princeton Plasma Physics Laboratory, Princeton, New Jersey 08540, USA}
\affiliation{Department of Physics, University of Maryland, College Park, Maryland 20742, USA}

\begin{abstract}
Strong magnetic fields are naturally self-generated in high-power, laser-solid interactions through the Biermann-battery mechanism. This work experimentally characterizes the 3D location and strength of these fields, rather than path-integrated quantities, through multi-view proton radiography and tomographic inversion on the OMEGA laser. We infer magnetic fields that extend several millimeters off the target surface into the hot, rarefied corona and are sufficient to strongly magnetize the plasma ($\Omega_{e}\tau_e \gg 1$). The data is used to validate MHD simulations incorporating recent improvements in magnetic transport modeling; we achieve reasonable agreement only with models with re-localization of transport by magnetic fields. This work provides a key demonstration of tomographic inversion in proton radiography, offering a valuable tool for investigating magnetic fields in laser-produced plasmas. 
\end{abstract}

\maketitle

Magnetic fields are commonly self-generated in high-power, laser-solid interactions through the Biermann-battery mechanism \cite{biermann_uber_1950,stamper_spontaneous_1971}. These fields enable laboratory studies of astrophysical processes including magnetogenesis \cite{gregori_generation_2015}, collisionless shocks \cite{campbell_formation_2024}, the Weibel instability \cite{fox_filamentation_2013}, magnetic reconnection \cite{nilson_bidirectional_2008,rosenberg_slowing_2015}, and astrophysical jets \cite{gao_mega-gauss_2019}. In addition, self-generated magnetic fields have been reported in simulations of National Ignition Facility (NIF) hohlraums where they affect the heat transport and implosion symmetry in inertial confinement fusion (ICF) and hohlraum experiments \cite{farmer_simulation_2017,walsh_self-generated_2017,leal_effect_2025}. Furthermore, externally applied magnetic fields have been shown to increase ion temperature and fusion yield in ICF implosion experiments \cite{moody_increased_2022, bailly-grandvaux_impact_2024}, highlighting the need to understand magnetic transport in high energy density conditions.

Prior characterization of self-generated fields has typically relied on proton radiography \cite{schaeffer_proton_2023} from a single line of sight, providing path-integrated measurements of magnetic fields through the plasma \cite{li_observation_2007,petrasso_lorentz_2009,willingale_fast_2010,gao_precision_2015,campbell_magnetic_2020,griff-mcmahon_measurements_2024}. Although some experiments have used two lines of sight \cite{li_measuring_2006} and a theoretical framework for multi-view proton imaging has been proposed \cite{spiers_methods_2021}, no study has measured the field distribution in the target-normal direction. As such, open questions remain about the structure and transport of self-generated magnetic fields.

\begin{figure} [b]
	\includegraphics[width=\linewidth]{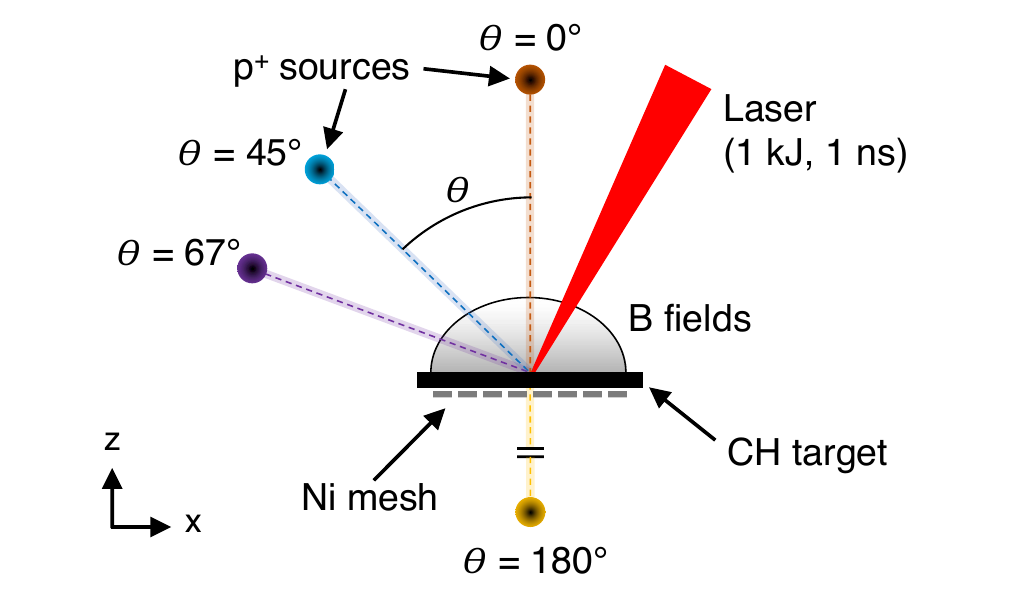}
	\caption{Experimental setup showing the different proton backlighter source positions (colored circles) in successive shots. The backlighter angle $\theta$ is defined relative to the target normal.}
	\label{fig:setup}
\end{figure}

\begin{figure*}
	\includegraphics[width=\linewidth]{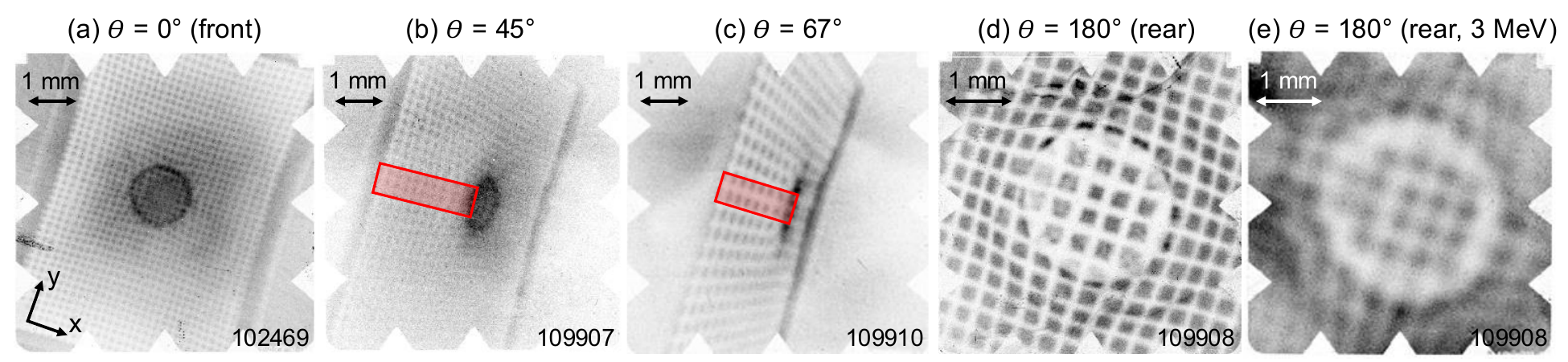}
	\caption{Experimental proton radiographs at $t=1.4~$ns from (a-d) 15 MeV protons and (e) 3 MeV protons, as the target is tilted about the $y$-axis in successive shots. Darker regions received higher proton fluence. The red boxes in (b,c) show where lineouts were taken in Figs. \ref{fig:deflection_comp}(d,e).}
	\label{fig:data}
\end{figure*}

Several extended magnetohydrodynamic (MHD) simulations have predicted that laser-ablated plasmas would become magnetized with electron Hall parameter $\Omega_e\tau_e \gg 1$, which has a significant effect on the electron temperature and global plasma evolution by inhibiting transport perpendicular to the magnetic field \cite{farmer_simulation_2017,walsh_kinetic_2024}. Meanwhile, other simulations have shown that the magnetic field is pushed out of the hot corona and anchored into the cold target by the Nernst effect (magnetic field advection by electron heat flux) \cite{nishiguchi_convective_1984}, leaving the coronal plasma weakly magnetized and with little effect on the transport \cite{lancia_topology_2014,gao_precision_2015,campbell_measuring_2022}. In general, simulations of self-generated magnetic fields have yielded conflicting results and predict markedly different field structure even for similar laser drive conditions \cite{li_measuring_2006,cecchetti_magnetic_2009,lancia_topology_2014,gao_precision_2015,campbell_measuring_2022}. These difference likely stem from the challenges in modeling laser-solid interactions, including steep spatial gradients at plasma-vacuum interfaces and uncertainties in how the Nernst effect should be treated when the heat transport is flux-limited \cite{walsh_kinetic_2024,brodrick_incorporating_2018,sherlock_suppression_2020}. Secondly, recent Vlasov-Fokker-Planck simulations have shown that Biermann-battery field generation is suppressed by nonlocal kinetic effects when the electron mean free path approaches the electron temperature gradient length scale ($\lambda_{ei}/L_T \gtrsim1$) \cite{sherlock_suppression_2020,davies_nonlocal_2023}, potentially further de-magnetizing laser-driven plasmas. Still other work has shown how magnetic fields can \textit{relocalize} transport \cite{froula_quenching_2007}, which would modify the effects just mentioned \cite{davies_nonlocal_2023} and give a complex picture of the interplay of magnetic fields and heat-flow in plasmas. Full 3D measurements of the magnetic field would resolve (1) the question of whether the coronal plasma is magnetized at all and (2) distinguish which extended MHD and kinetic effects need to be included in laser-solid interaction models.

In this Letter, we characterize the 3D structure of self-generated magnetic fields in a high power laser-solid interaction. A multi-view proton radiography scheme is used to image the fields from four viewpoints in successive laser shots, enabling a highly-constrained tomographic inversion. We measure magnetic fields that extend several millimeters away from the target into the corona with a 1.8 T volume-averaged field, sufficient to strongly magnetize the plasma (volume-averaged $\Omega_e\tau_e \approx 1000$). We then validate extended MHD simulations that produce similar coronal magnetic fields (within 50\%), but interestingly only when Biermann-battery field generation is at full strength (i.e. not suppressed by kinetic effects). The field structure suggests that laser-ablated plasmas are strongly magnetized, which has important implications for heat transport relevant to hohlraum physics and laser-heated plasmas in general.

The experiment was conducted on the OMEGA laser at the Laboratory for Laser Energetics. A 25-$\mu$m thick CH foil was irradiated by two overlapped laser beams, which produced self-generated magnetic fields in the ablated plasma plume. The laser beams delivered 1 kJ of energy in 1 ns over a $1/e$ laser radius of 358 $\mu$m using SG5 Distributed Phase Plates \cite{lin_distributed_1995}, resulting in a peak laser intensity of $3\times10^{14}~$W/cm$^2$. A separate set of 25 beams were used to implode a D$^3$He gas-filled capsule to produce a backlighter source of 3 and 15 MeV protons and x-rays that imaged the magnetic fields from the target interaction at $t=1.4$ ns after the drive beams turned on. A Ni mesh was attached to the rear side of the target to perform mesh proton radiography with x-ray fiducials \cite{johnson_proton_2022,malko_design_2022}. Four different backlighter capsule positions were used in successive shots to image the magnetic fields from the laser-solid interaction, as shown in Fig. \ref{fig:setup}. The backlighters were positioned 10 mm from the target and the laser angle of incidence and other laser parameters were held constant throughout the experiment by firing different beams as the target was rotated to produce similar fields for each shot. The target tilt was known to within a few degrees for each shot.

High contrast proton radiographs were acquired from the four different viewing angles, shown in Fig. \ref{fig:data}. Each radiograph has a corresponding x-ray image (see Appendix) that acts as a reference for the unperturbed mesh \cite{johnson_proton_2022,malko_design_2022,griff-mcmahon_measurements_2024,griff-mcmahon_proton_2024}. For each mesh beamlet, the absolute proton deflection is measured by comparing the mesh position between the proton and x-ray images. The deflection encodes information about the path-integrated magnetic fields along the proton trajectory. The different view angles break the degeneracy of the path-integrated diagnostic and are used to recover the three-dimensional field structure through a quantitative tomographic inversion. More directly, there is qualitative evidence that these images contain tomographic information about the magnetic fields from the ``Biermann ring", which is the dark central feature in Fig. \ref{fig:data}(a-c) where protons are focused onto each other. We observe an apparent shift of this feature from the center of the foil in the oblique views [Fig. \ref{fig:data}(b,c)], indicating that the fields are located some height above the target surface. 3 MeV protons emitted from the same proton source were also measured in the rear-view case [Fig. \ref{fig:data}(e)], and provide additional information about the relative contribution of electric fields.

\begin{figure}
	\includegraphics[width=\linewidth]{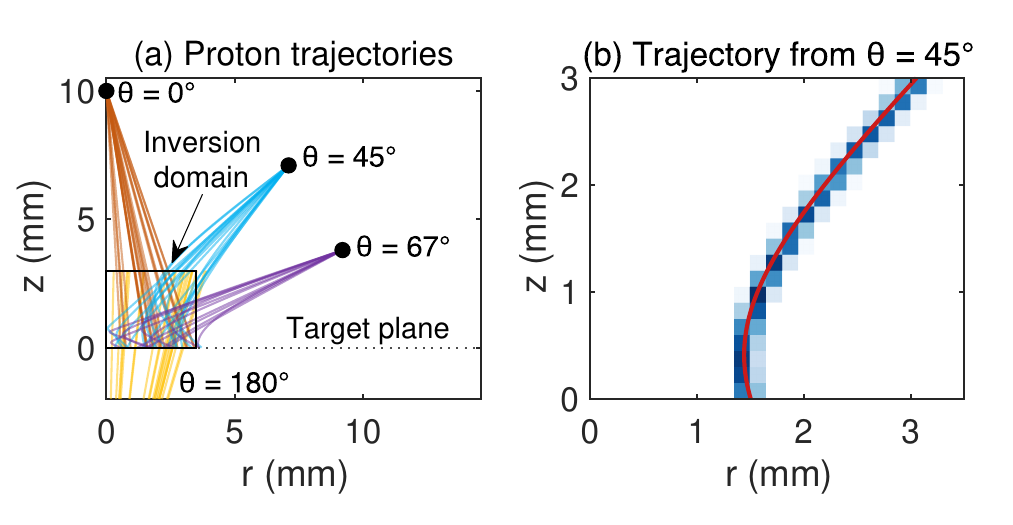}
	\caption{(a) Sample straight-line proton trajectories from the different backlighter sources, projected onto cylindrical coordinates. The target is positioned on the $z=0$ plane and irradiated from the positive $z$ direction. The black box shows the inversion domain. (b) Zoomed-in view of a single proton trajectory from the $\theta=45\degree$ backlighter through the inversion domain. The trajectory is discretized on a 2D grid with 150$~\mu$m resolution and the pathlength in each cell is shown by the blue color. The red line shows the proton trajectory.}
	\label{fig:RZ_rays}
\end{figure}

Several simplifying assumptions are imposed to aid in the tomographic inversion procedure. First, we adopt an axisymmetric model of the toroidal Biermann-battery magnetic field $B_\phi(r,z)$ so that the 3-dimensional proton trajectories are collapsed onto the 2D $rz$-plane. This increases the tomographic coverage over the domain and makes the inversion tractable. Figure \ref{fig:RZ_rays}(a) shows a subset of proton trajectories from each backlighter through the $rz$-space. Information about the field structure is extracted where proton paths cross each other in this figure. 
The second assumption is that proton deflections are perturbative so that the protons sample the fields exactly along the x-ray trajectories, which are straight lines. Projection of these trajectories into cylindrical coordinates leads to the curved paths shown in Fig \ref{fig:RZ_rays}(a). This assumption is validated after the inversion. Over 2000 mesh beamlets were tracked in total across the different shots to recover vector deflection information about the path-integrated fields for each proton trajectory.

The algebraic reconstruction tomography (ART) framework is used to tomographically invert the proton radiographs \cite{kak_principles_2001}. This framework discretizes the fields in a domain and solves a system of linear equations based on the path-integrated deflection data. ART is typically performed with scalar quantities like density, but here it was adapted to deal with vector fields. Using this method, proton deflections measured on the detector $\vec d$ are approximated as a series of deflection contributions along the proton trajectory:
% \begin{equation} \label{eq:deflection}
    $\vec d = \frac{e L}{m_p v_p^2} \sum_k dl_k\  \vec v_p \times \vec B_k.$
% \end{equation}
Here, $e$ is electron charge, $L$ is the distance from the target to the detector, $m_p$ is proton mass, $v_p$ is proton velocity, and the $\perp$ symbol references the field component perpendicular to the proton trajectory. The index $k$ represents the $k^{th}$ pixel in the domain and applies to the proton pathlength $dl_k$ and the magnetic fields $\vec B_k$. Figure \ref{fig:RZ_rays}(b) shows an example of a single proton trajectory through the domain where the total path-length in each cell $dl_{k}$ is shown in blue. A linear system of equations is constructed as $\mathbf{A}\vec x=\vec b$ where $\mathbf{A}$ is a matrix containing information about the ray trajectory and geometric factors, $\vec x$ describes the magnetic fields, and $\vec b$ contains the measured deflections on the detector for each beamlet. 
The system is iteratively solved using a least squares solver. Additional information about the inversion process is found in the Appendix.

\begin{figure}
	\includegraphics[width=\linewidth]{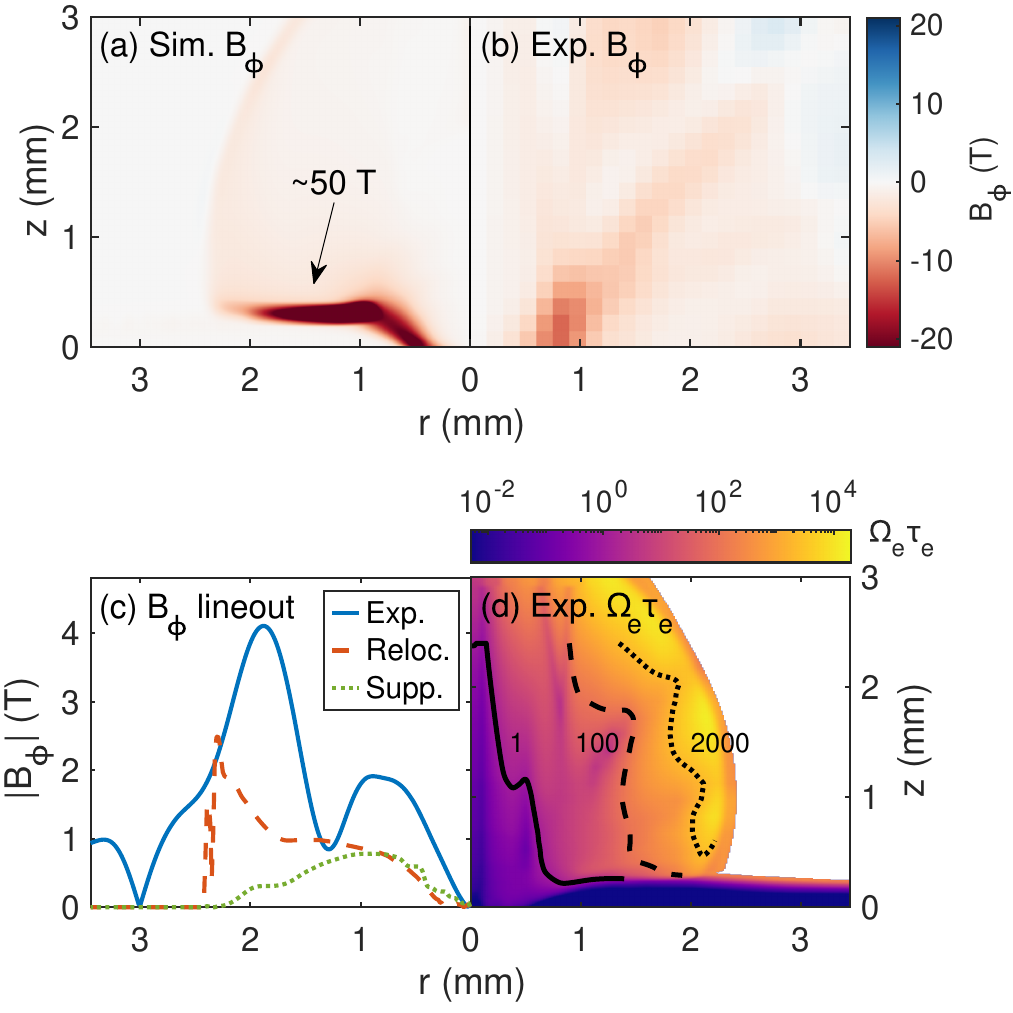}
	\caption{(a) Toroidal magnetic field from extended MHD simulation at full field strength and (b) extracted from the tomographic inversion. All fields have been smoothed over 100 $\mu$m and the color limits have been adjusted to compare the coronal fields. (c) Radial lineout of $|B_\phi|$ at $z=1.5~$mm for experiment, relocalized simulation, and maximally suppressed simulation using a recent model of Biermann suppression by nonlocal effects \cite{davies_nonlocal_2023}. (d) Electron hall parameter estimated from experimental magnetic fields and simulated density and temperature. The solid, dashed, and dotted lines are isocontours of $\Omega_e \tau_e=$ (1, 100, 2000), respectively. }
	\label{fig:B_inversion_GORGON}
\end{figure}

Figure \ref{fig:B_inversion_GORGON} shows the toroidal magnetic field extracted from the tomographic inversion and compared to an extended MHD simulation. The experimental fields extend out into the coronal plasma and contrast sharply with prior simulations from Li \emph{et al.} \cite{li_measuring_2006,li_observation_2007} that reported magnetic fields confined to a thin shell at the plasma bubble edge, as well as with more recent simulations \cite{lancia_topology_2014,gao_precision_2015} that showed magnetic fields anchored close to the target surface. The simulation in Fig. \ref{fig:B_inversion_GORGON}(a) used the \textsc{gorgon} code in 2D cylindrical coordinates \cite{ciardi_evolution_2007,chittenden_recent_2009,walsh_extended-magnetohydrodynamics_2020} and contained Biermann-battery field generation, Nernst advection, radiation transport, and improved magnetic transport coefficients at low magnetization \cite{davies_transport_2021,sadler_symmetric_2021,walsh_updated_2021}. 
This simulation included a recent model for Biermann-battery suppression with relocalization that turns on for magnetized plasmas from Ref. \cite{davies_nonlocal_2023}. Interestingly, the relocalization is strong enough that the fields are nearly equivalent to a simulation that includes the full Biermann effect without any suppression.
The strongest magnetic fields in both the inversion and simulation are located near the edge of the laser spot at $r\approx1$ mm and within 0.5 mm above the target surface. In simulations, this is where the electron density and temperature gradients are the greatest and where the Biermann-battery source term is the largest $(\partial B/\partial t = \nabla T_e \times \nabla n_e / e n_e)$. It also coincides with the interface between the cold target and hot corona, where Nernst advection anchors most of the simulated fields to the target. In addition to the fields close to the target, both the simulation and experiment show magnetic fields extending several millimeters off the target into the corona at the $\sim1~$T level.

We also compared the results to a simulation with maximal suppressive effects that does not account for relocalization \cite{davies_nonlocal_2023}. Figure \ref{fig:B_inversion_GORGON}(c) compares radial lineouts of $|B_\phi|$ along $z=1.5~$mm for experimental fields, a simulation with relocalization, and a simulation with maximal Biermann suppression \cite{davies_nonlocal_2023}. Interestingly, we see best agreement for the simulation that includes relocalization. The volume-averaged coronal field strength (defined for $z>1~$mm) is $1.8~$T for experiment, $1.0~$T for relocalized simulation, and $0.3~$T for maximally suppressed simulation. The relocalized simulation obtains a similar magnetic field as the data (within a factor of 2), however including the maximal Biermann suppression effects significantly underpredicts the coronal fields by a factor of more than 5. We also evaluated the total magnetic flux $\iint B_\phi\,dr\,dz$ and found the maximally suppressed simulation underpredicts the flux by more than a factor of 2; the experiment, relocalized simulation, and maximally suppressed simulation have magnetic fluxes of 21, 17, and 9 T$~$mm$^2$, respectively.

The physical implications of extended magnetic fields in the corona can be described by the degree of electron magnetization, denoted by the Hall parameter $\Omega_e \tau_e$, shown in Fig. \ref{fig:B_inversion_GORGON}(d). Here, $\Omega_e$ is the electron cyclotron frequency and $\tau_e$ is the electron-ion collision time. The Hall parameter gives the average number of gyro-orbits an electron will make between collisions and scales as $\Omega_e \tau_e \sim B T_e^{3/2} / n_e$ so that even relatively weak fields can significantly magnetize the hot and rarefied corona. The volume-averaged Hall parameter in the corona ($z>1~$mm) is $\sim1000$ for experiment, $\sim500$ for clean simulation, and $\sim100$ for suppressed simulation. Auxiliary experiments support using simulated density and experimental fields to estimate the experimental Hall parameter \cite{mccluskey_reconstruction_2025}. We will return to a detailed discussion of the implications for MHD simulations in the discussion below.

\begin{figure}
	\includegraphics[width=\linewidth]{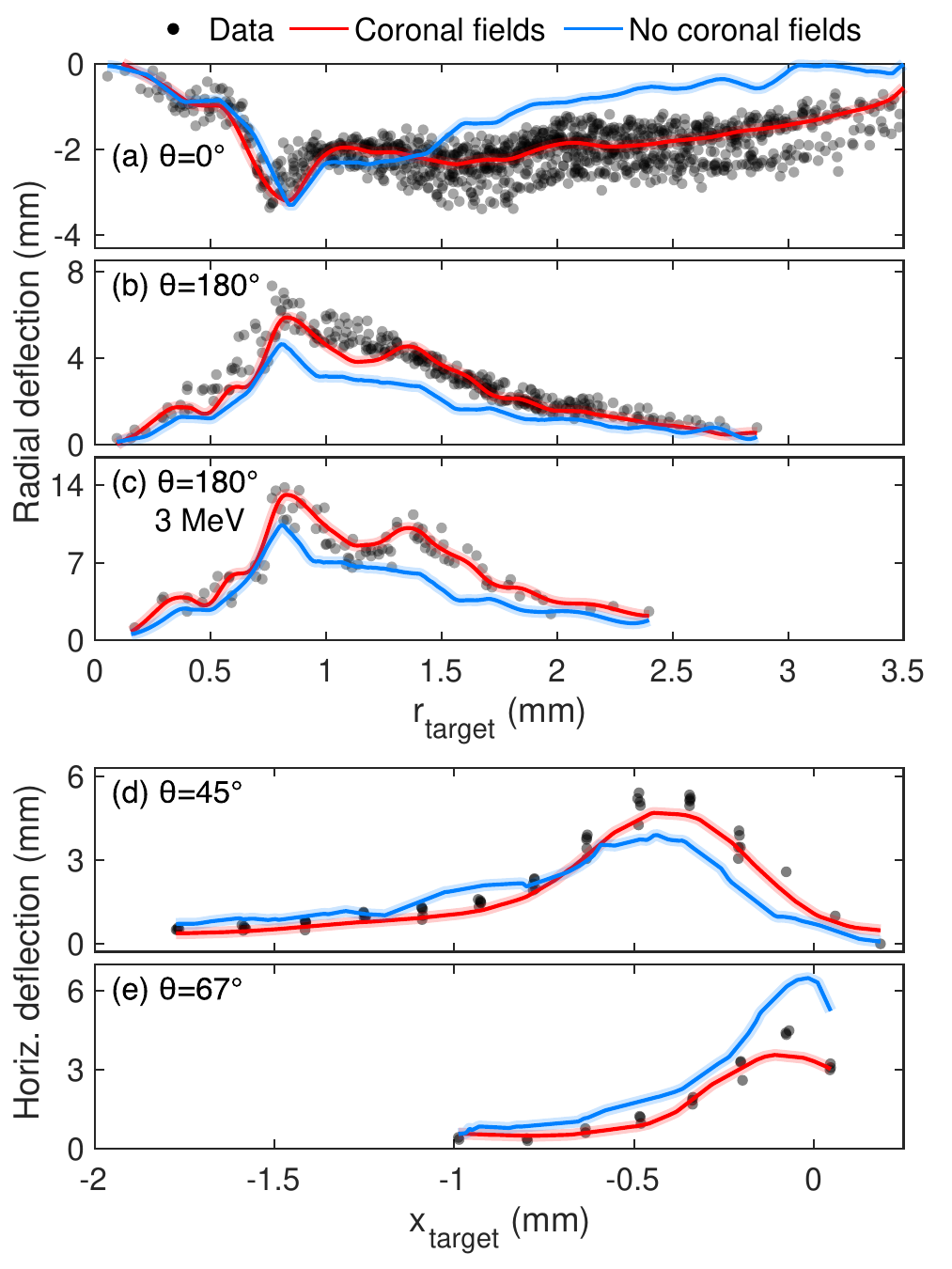}
	\caption{Lineouts of the proton deflection on the detector in the (a-c) radial direction for front and back views and in the (d,e) horizontal direction along the x-axis for $\theta=45\degree$ and $67\degree$ views from the regions outlined in Fig. \ref{fig:data}. The experimental data (black circles) are compared to synthetic deflection from inversions that include coronal fields (red lines) and omit coronal fields (blue lines).}
	\label{fig:deflection_comp}
\end{figure}

Given the significant differences in the simulated magnetic fields in the corona between different Biermann suppression models [Fig. \ref{fig:B_inversion_GORGON}(c)], we now carefully confirm that the data support appreciable magnetic fields in the corona. To do so, we conducted an additional tomographic inversion for comparison, in which all magnetic fields for $z>0.5~$mm were kept at zero, which we call the ``no coronal fields" inversion. Figure \ref{fig:deflection_comp} compares experimental deflection data and synthetic proton deflections that were generated by forward modeling proton trajectories through both inversions. In the inversion with coronal fields, synthetic deflections (red lines in Fig. \ref{fig:deflection_comp}) follow closely with the experimental data (black dots) for all views. The deflection residual error is 21\% and is dominated by azimuthal variation in the deflections that cannot be captured by an axisymmetric model and manifest as a scatter of the data in Fig. \ref{fig:deflection_comp}. In contrast, the inversion without coronal fields is poor and has a 44\% normalized error and significant apparent systematic differences. There are large regions where the ``no corona" synthetic deflections (blue lines) deviate from the experimental data. For example, this inversion cannot simultaneously satisfy the proton deflections at large radius; the front and back views [Fig. \ref{fig:deflection_comp}(a,b)] measure strong deflections while the angled views [Fig. \ref{fig:deflection_comp}(d,e)] measure relatively weak deflections. Consequently, the measurements demand substantial magnetic fields at $z>0.5$ mm into the corona.

In addition to statements about the domain size, a comparison between the 3 and 15 MeV proton deflections yields information about the radial electric field. Magnetic deflection scales with proton energy as $d_B \sim E_p^{-1/2}$ while electric deflection scales as $d_E \sim E_p^{-1}$. In the rear view, the deflection ratio between the two proton populations is consistent with a purely magnetic deflection. This statement is further supported by the agreement of synthetic deflection data for both energies from an inversion that only includes magnetic field [red lines in Fig. \ref{fig:deflection_comp}(b,c)]. Furthermore, including a radial and/or z-directed electric field into the inversion itself does not change the conclusion about the magnetic fields extending off the target or significantly modify line-averaged quantities like $\int B\, dr$ or $\int B\, dz$. However, the electric fields are more difficult to constrain due to the subdominant nature of electric deflections for 15 MeV protons and likely require an additional investigation at lower proton energy.

We now discuss the implications for the observations of strong coronal magnetic fields and our initial comparisons against various extended MHD simulations. The first important conclusion is that we infer strong magnetic fields throughout the corona, which are sufficient to substantially magnetize the plasma (Hall parameter $\Omega_e \tau_e \gg 1)$. The Hall parameter impacts several magnetized transport processes like the electron thermal conductivity (proportional to temperature-gradient heat flux), which is suppressed perpendicular to the magnetic field by a factor $\kappa_\perp/\kappa_\parallel \sim  1/ (1+\Omega_e^2\tau_e^2)$ relative to the unmagnetized case \cite{braginskii_transport_1965}. Modification of the heat flux directly impacts the global plasma evolution through changes to the electron temperature \cite{farmer_simulation_2017,lezhnin_simulations_2025}. Magnetization of the coronal plasma would have broad impacts for indirect-drive hohlraum experiments and ICF, where plasma transport off the the hohlraum walls and laser entrance hole is crucially important for laser-plasma coupling and influences x-ray radiation drive, implosion symmetry, and laser-plasma instabilities \cite{farmer_simulation_2017,walsh_kinetic_2024,leal_effect_2025}. Additionally, extended magnetic fields are relevant for laboratory astrophysical experiments and could affect the physical mechanisms and interpretation of collisionless shock \cite{campbell_formation_2024}, the Weibel instability \cite{fox_filamentation_2013}, and magnetic reconnection experiments \cite{nilson_bidirectional_2008,rosenberg_slowing_2015}. We note that several prior works have cast doubt on whether significant coronal fields could exist, either through the effect of Nernst advection of magnetic fields into the solid target \cite{lancia_topology_2014,gao_precision_2015}, or through effects suppressing the Biermann battery \cite{sherlock_suppression_2020,davies_nonlocal_2023}. This work shows a clear case with relevant laser intensity and geometry with a significant coronal field and Hall parameter much greater than 1.

Second, our results are consistent with minimal suppression by nonlocal effects. Often, the coronal plasma is treated as nonlocal ($\lambda_{ei}/L_T \gtrsim1$) \cite{henchen_observation_2018}, which is theorized to suppress the Biermann-battery field generation \cite{sherlock_suppression_2020,davies_nonlocal_2023}. However, if the plasma becomes magnetized ($\Omega_e \tau_e > 1$), the relevant transport length scale shrinks from the electron mean free path to the Larmor radius, as observed in \cite{froula_quenching_2007}. In this case, the nonlocal suppression of the Biermann-battery effect is reduced due to relocalization. This is contained in the relocalized Biermann model \cite{davies_nonlocal_2023}, which provides the results most consistent with experimental magnetic fields. Strong magnetization also inhibits Nernst advection and would lead to radically different magnetic field dynamics. Although the Biermann term may be suppressed near the target where the transport is nonlocal and unmagnetized, it should remain at full strength in the corona due to strong magnetization and re-localization. Despite the agreement in coronal field strength between experiment and clean simulation, there remains a large discrepancy in field strength close to the target; the fields in the simulation are compressed and anchored into a pancaked region at the target interface, whereas the experiment shows no such anchoring feature. It is clear that the current models for Biermann-battery and magnetic transport require additional development beyond the scope of this work to further match other aspects of the magnetic field structure. 

In conclusion, we have reconstructed the three-dimensional structure of self-generated magnetic fields from a laser-solid interaction using tomographic proton radiography. The fields extend several millimeters off the target surface into the low density, high temperature corona, magnetizing the plasma to Hall parameter $\Omega_e\tau_e \gg 1$. Extended MHD simulations reproduce similar coronal field strengths ($\sim1~$T) when the Biermann-battery effect is at full strength and unsuppressed from nonlocal effects.
Coronal fields of this magnitude will significantly modify plasma transport in hohlraum environments \cite{farmer_simulation_2017,walsh_kinetic_2024} and calls for additional investigation of magnetic field generation and transport.
Finally, this work is a first demonstration of tomographic proton radiography and establishes a powerful tool for probing magnetic fields in laser-produced plasmas.

\section*{Acknowledgements}
The authors thank the LLE staff for their help in conducting these experiments and the MIT team and Melody Scott, Peter Heuer, and Hannah McClow at LLE for processing the CR39 data. 

The work was performed under the auspices of the U.S. Department of Energy by General Atomics under NNSA Contract 89233124CNA000365 and by Lawrence Livermore National Laboratory under Contract DE-AC52-07NA27344. The experiment was conducted at the Omega Laser Facility with beam time through the National Laser Users’ Facility user program. This work was supported by the Department of Energy under grant Nos. DE-NA0004034, DE-NA0004271, and DE-NA0003868 and by the University of Rochester ``National Inertial Confinement Fusion Program" under award No. DE-NA0004144. This work is supported by the National Science Foundation Graduate Research Fellowship Program under grant No. 2039656.

\appendix

\section{Tomographic Validation}

\begin{figure} [b]
	\includegraphics[width=0.9\linewidth]{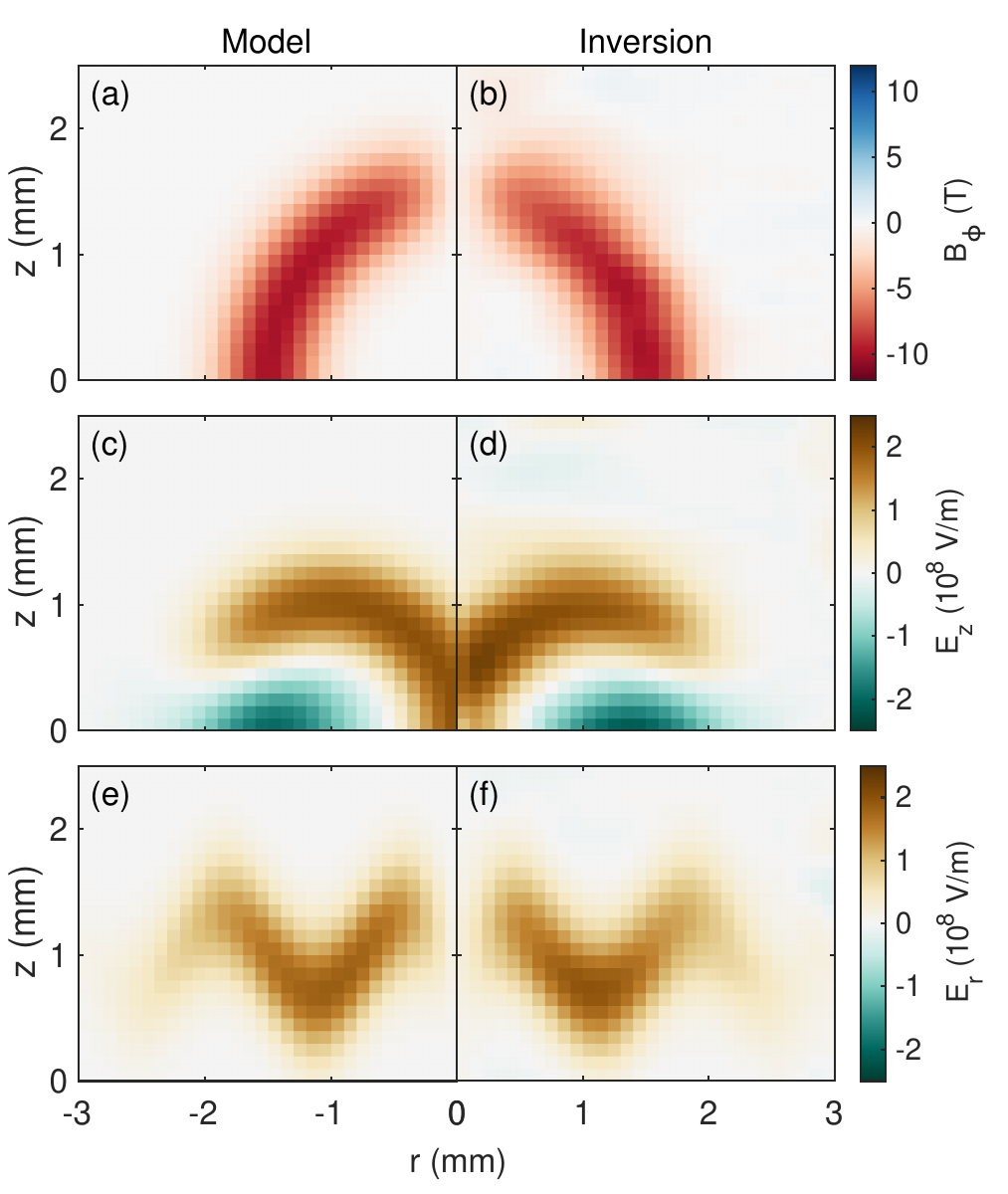}
	\caption{Comparison of model fields (left) and inverted fields using synthetic deflection maps from the four experimental view angles (right). All three field components ($B_\phi$, $E_z$, and $E_r$) were included to generate the deflection maps and were decoupled in the inversion.}
	\label{fig:TomoValidation}
\end{figure}

The tomographic inversion procedure was validated using synthetic proton deflection maps from each of the four experimental view angles. Three axisymmetric fields, $B_\phi$, $E_z$, and $E_r$, were included to generate the deflection maps, with profiles shown in the left column of Fig. \ref{fig:TomoValidation}. The magnetic profile is a hemisphere shell centered at the origin, the z-directed electric field is part of a toroidal shell with a central feature of the opposite sign, and the radial electric field is an undulating line of field located at a height of $z=1$ mm above the target. These field strengths and profiles were chosen to test possible failure modes of the experimental inversion and investigate the effects of fields with large radius, large height, sign reversals, and overlap between the different components.

The right column of Fig. \ref{fig:TomoValidation} shows the tomographically inverted fields. Excellent agreement is observed for all field components with clear distinctions between regions where fields were imposed and the surrounding regions without fields. The inversion was also able to isolate the electric and magnetic fields despite significant spatial overlap. The root mean squared error between the model and inversion fields is 0.37 T, $1.1\times10^7$ V/m, and $6.2\times10^6$ V/m for $B_\phi$, $E_z$, and $E_r$, respectively. These correspond to relative errors of 3.7\%, 5.7\%, and 3.1\%, compared to the characteristic field strengths in each profile. These small errors are further reduced when weaker fields are used; protons are assumed to sample the fields along straight-line trajectories, which breaks down when the fields are sufficiently strong and extended along the proton path. For 15 MeV protons, this assumption is valid for proton deflections $\lesssim3\degree$ and extend $\lesssim$ 2 mm. Future work may use an iterative scheme that self-consistently updates the proton trajectories based on the inverted fields.

\section{Tomographic Setup} \label{app:tomo_def}
\begin{figure}
	\includegraphics[width=0.9\linewidth]{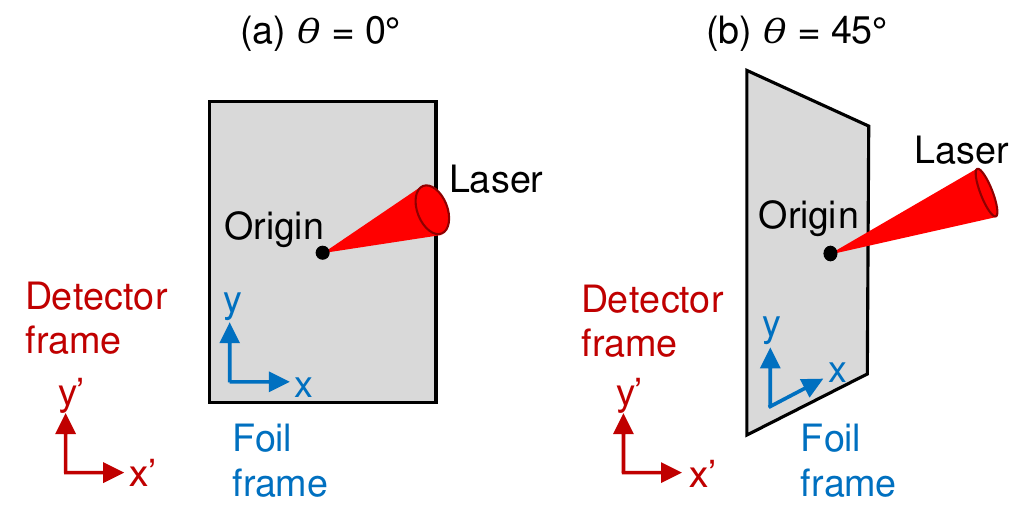}
	\caption{Foil target viewed along the backlighter axis for tilt angles of (a) $\theta=0\degree$ and (b) $\theta=45\degree$. The detector coordinate frame is shown in red with primed coordinates and the foil coordinate frame in blue with unprimed coordinates.}
	\label{fig:coord_sys_tomo}
\end{figure}

In order to connect the different tomographic views, we define a common coordinate system in the frame of the foil target ($x,y,z$) as shown in Fig. \ref{fig:coord_sys_tomo}. The $xy$-plane lies along the foil. A detector coordinate system with primed coordinates ($x',y',z'$) is also defined, where the $z'$-axis is aligned with the proton backlighter axis. Both systems share a common origin at the center of the target and are related through a rotation along the y-axis by the backlighter angle $\theta$. 

The angular deflection of a proton traversing through the field region is 
\begin{equation}
    \vec\alpha = \frac{e}{m_p v_p^2}\int dl\, \left[\vec E_\perp + \vec v_p \times \vec B \right]
\end{equation}
where the $\perp$ symbol references the field component perpendicular to the proton trajectory. The angular deflection $\vec \alpha$ can be broken into components $\alpha _{r'}$ for deflections in the $\hat{\phi}'\times\hat{v}_p$ direction (equivalent to $\hat{r}'$ in the paraxial limit) and $\alpha_{\phi'}$ for deflections in the $\hat{\phi}'$ direction, which is the azimuthal coordinate in the detector frame.

Without assumptions of paraxiality, the deflection $d$ on the detector is
\begin{align}
    d_{r'}&=L \alpha_{r'} \sec^2{\theta'}\\
    d_{\phi'}&=L \alpha_{\phi'} \sec{\theta'}
\end{align}
where $L$ is the distance from the field to the detector, $\theta'$ is the polar angle from target normal, and the $\sec{\theta'}$ factor(s) are non-paraxial corrections that account for increased path-length between the fields and the detector and for geometric projection on the detector. The radial and azimuthal deflections are used rather than the standard Cartesian to allow for easy extension to the non-paraxial limit, which gives corrections up to the 15\% level based on the experiment field of view.

The proton source is located on the $z'$-axis so that proton velocities have only $\hat z'$ and $\hat r'$ components. Therefore, the radial and azimuthal deflections in the detector frame are given below.
\begin{align}
    \alpha_{r'}&= \frac{e}{m_p v_p^2} \int dl \, \left[v_p B_{\phi'} + E_{r'}\cos \theta' + E_{z'}\sin \theta' \right]\\ \label{eqn:alpha_r}
    \alpha_{\phi'}&= \frac{e}{m_p v_p^2} \int dl \, \left[v_p (B_{r'} \cos \theta' - B_z' \sin\theta') + E_{\phi'}\right]
\end{align}

Since the fields are defined in the foil frame, we must calculate the transformation from the foil frame to detector frame using the transfer matrix $\mathbf{T}$.
\begin{equation}
\begin{bmatrix}
B_{r'}\\
B_{\phi'}\\
B_{z'}
\end{bmatrix}
=
\mathbf{T}
\begin{bmatrix}
B_{r}\\
B_{\phi}\\
B_{z}
\end{bmatrix}\label{eqn:CoordTransToUnprimed}
\end{equation}

\begin{equation}
\mathbf{T} = \mathbf{C}\ \mathbf{R_y(\theta)}\  \mathbf{D} \label{eqn:Transfer_def}
\end{equation}

\begin{equation}
\mathbf{C}=
\begin{bmatrix}
\cos \phi' & \sin \phi' & 0\\
-\sin\phi' & \cos\phi' & 0\\
0 & 0 & 1
\end{bmatrix}
,\ 
\mathbf{D} = \begin{bmatrix}
\cos \phi & -\sin \phi & 0\\
\sin\phi & \cos\phi & 0\\
0 & 0 & 1
\end{bmatrix}
\end{equation}
Here, $\mathbf{D}$ converts the foil fields from cylindrical to Cartesian in the foil frame, $\mathbf{R_y(\theta)}$ rotates the fields to the detector frame and $\mathbf{C}$ converts the detector fields back to cylindrical coordinates in the detector frame.

\section{Geometric Corrections for Mesh}
\begin{figure}
	\includegraphics[width=0.9\linewidth]{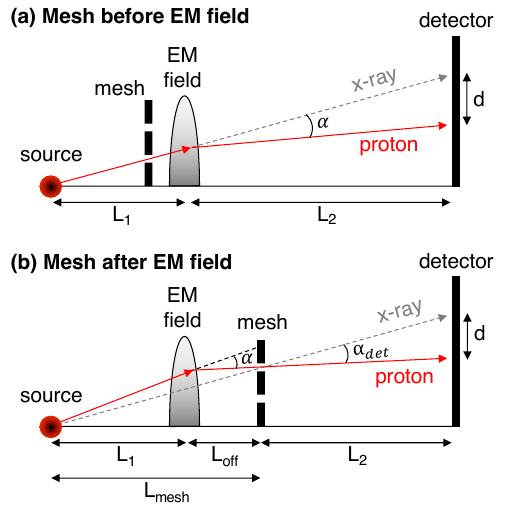}
	\caption{Proton radiography setups with mesh placed (a) between source and EM fields and (b) between EM fields and detector. In the latter case, a correction is needed to relate the detector deflection $d$ to the true deflection angle $\alpha$.}
	\label{fig:MeshSetup}
\end{figure}

In mesh proton radiography, the mesh is typically placed between the proton source and the electromagnetic (EM) fields as shown in Fig. \ref{fig:MeshSetup}(a). The mesh splits the x-rays and protons into beamlets that can be tracked on the detector. X-rays serve as a reference for the unperturbed proton trajectory. In the paraxial and small deflection-angle approximations, the proton deflection on the detector $d$ is proportional to the deflection angle $\alpha$ and field-to-detector distance $L_2$ according to $d=L_2 \alpha$. 

However, the shots discussed in the main text use an alternative scheme where the mesh is attached to the rear surface of the target. In some laser configurations, the protons encounter the fields before passing through the mesh [Fig. \ref{fig:MeshSetup}(b)]. This setup efficiently incorporates a mesh into the experiment without requiring an additional OMEGA Ten-Inch Manipulator (TIM). However, the proton deflection analysis is more complicated since protons and x-rays may take different initial trajectories before passing through the same mesh hole. In this case, the proton deflection on the detector is related to the exit angle through the mesh $\alpha_{det}$ where $\alpha_{det} \neq \alpha$. This requires a correction to relate the proton deflection on the detector $d$ with the proton deflection angle $\alpha$.

We consider a thin field region located at some offset distance $L_{off}$ before the mesh [Fig. \ref{fig:MeshSetup}(b)]. Under the assumption of small angle deflections ($\alpha \ll 1$ rad), the correction to the deflection angle is geometrically inferred as
\begin{equation}
    \alpha_{det} = \alpha\left(1 - \frac{L_{off}}{L_{mesh}} \right) \label{eqn:meshCorr1}
\end{equation}
where $\alpha_{det}$ is the deflection angle measured on the detector, $\alpha$ is the true deflection angle that occurs in the field region, and $L_{mesh}$ is the distance between the proton source and mesh. Therefore, a field located at a distance of 2 mm off the target, and a mesh distance of 10 mm yields an underestimate of the true deflection angle by 20\%.

If there are multiple deflections at different distances before the mesh, Eq. \eqref{eqn:meshCorr1} extends to
\begin{equation}
    \alpha_{det} = \sum_i\alpha_{i}\left(1 - \frac{L_{off,i}}{L_{mesh}} \right) \label{eqn:meshCorrAll}
\end{equation}
where $\alpha_i$ is the $i^{th}$ field deflection at a distance $L_{off,i}$ before the mesh.  Eq. \eqref{eqn:meshCorrAll} can now be implemented into the tomography scheme to account for geometries where the proton passes through the field region first, and then through the mesh.

\section{X-ray Reference Images}
\begin{figure} [h!]
	\includegraphics[width=\linewidth]{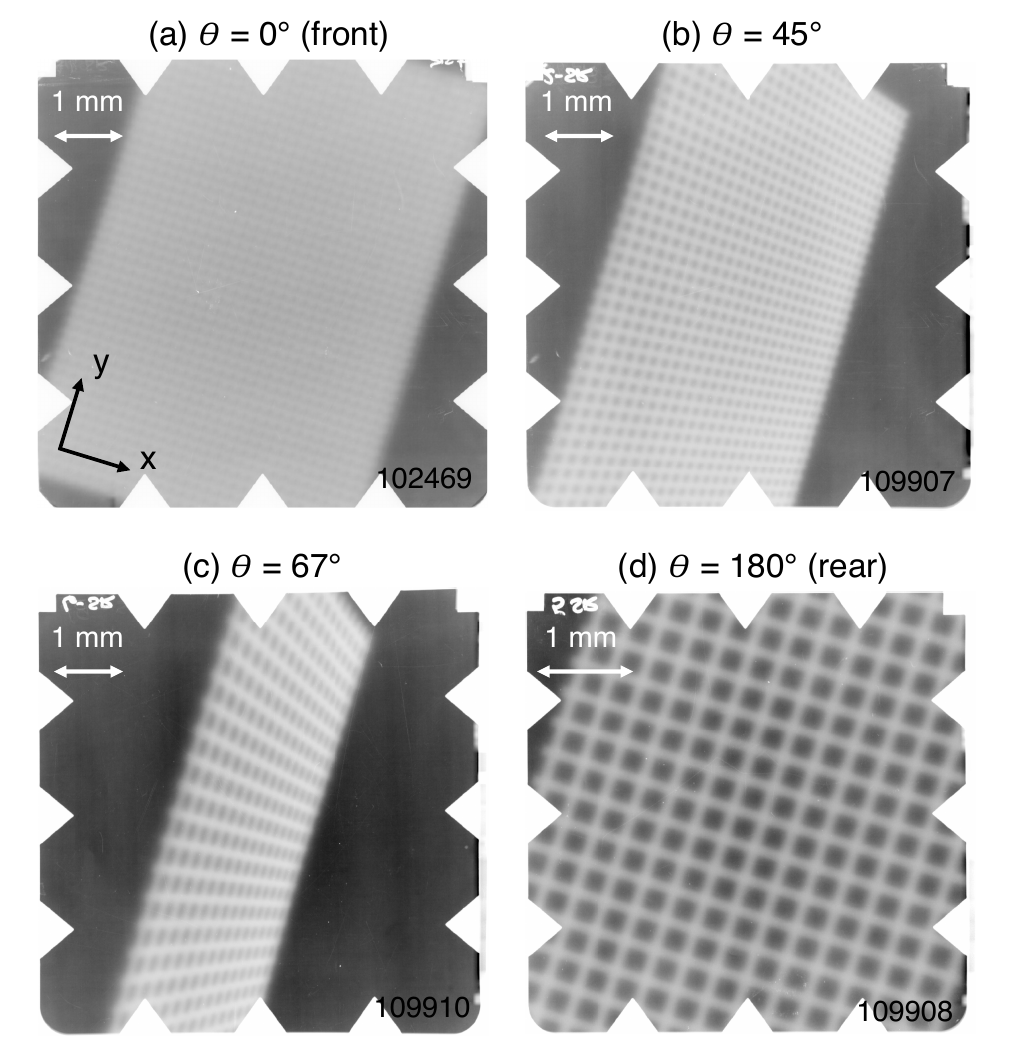}
	\caption{X-ray image plate data used to recover the unperturbed proton locations for each view angle. These images were used in tandem with Fig. \ref{fig:data} in the main text to determine the proton deflections.}
	\label{fig:IP_data_4}
\end{figure}

X-ray images of the mesh target were acquired by placing an image plate at the rear of the detector stack and shown in Fig. \ref{fig:IP_data_4}. Each image was aligned to its corresponding proton image from Fig. \ref{fig:data} in the main text, using the jagged fiducials along the boundary. Once aligned, the mesh in the x-ray image gave the location of unperturbed protons and was used to determine the absolute proton deflection by comparing x-ray and proton mesh beamlets. 

\bibliography{Prad_Tomo}% Produces the bibliography via BibTeX.

\end{document}